# On the use of naked-eye sunspot observations during the Maunder Minimum

V.M.S. Carrasco[1,2,*] • M.C. Gallego[1,2] • R. Arlt[3] • J.M. Vaquero[2,4]

[1] Departamento de Física, Universidad de Extremadura, 06006 Badajoz, Spain

[2] Instituto Universitario de Investigación del Agua, Cambio Climático y Sostenibilidad (IACYS), Universidad de Extremadura, 06006 Badajoz, Spain

[3] Leibniz Institute for Astrophysics Potsdam, An der Sternwarte 16, 14482 Potsdam, Germany

[4] Departamento de Física, Universidad de Extremadura, 06800 Mérida, Spain

* Corresponding author: vmscarrasco@unex.es

**Abstract:** Naked-eye sunspot observations (NESO, hereafter) have been recorded for last two millennia, approximately. This kind of records were made around the world, mainly in Asian civilizations, and they are compiled in several catalogues. In this work, we analyze solar activity in days of the 19th century when NESO were recorded. We found that only more than five sunspot groups were recorded in 39 % of days corresponding to these NESO events. Furthermore, regarding the largest groups observed in days when NESO were reported, we show the uncorrected areas of these groups were below 200 millionths of solar disc (msd hereafter) in 3.2 % of total cases while it is 12.9 % for areas between 200 and 499 msd. Thus, NESO records do not imply high solar activity and big sunspot groups necessarily. Therefore, these results contradict the interpretations of recent works that, using the same NESO set, suggest the solar activity level during the Maunder Minimum is still an open question. NESO records support the Maunder Minimum as a very low solar activity period.

**Unified Astronomy Thesaurus concepts:** Sunspots (1653); Solar activity (1475); Maunder minimum (1015)

**1. Introduction**

Sunspots can be observed with the naked eye under certain conditions of visibility (Schaefer 1993). NESO constitute a heritage of enormous interest for the history of astronomy and the study of the physical conditions of the Sun during the last twenty centuries (Vaquero & Vázquez 2009). In recent decades, several researchers have made a great effort to understand the advantages and limitations of this type of records. Thus, several studies have concluded that historical, social and meteorological factors clearly affect the record frequency of these observations (Hameed & Gong 1991, Chen et al. 2020).

Most of NESO were made by Asian civilizations such as China, Korea, and Japan (Keimatsu 1970, Tamazawa et al. 2017), although there are also western and other records around the world (Willis & Stephenson 2001, Vaquero & Gallego 2002). We can note many observations made by Asian astronomers were carried out during sunsets in order to try to detect the first quarter of the Moon which regulated the calendar. There are around 300 known NESO preserved in historical documents and collected in different catalogues (Wittmann & Xu 1987, Yau & Stephenson, 1988). This observation set has been used in different studies of solar activity. For example, Vaquero et al. (2002) presented an annual series of NESO for the period 165 BC–1918 AD showing a 250-year cycle and Lee et al. (2004) studied the sunspot and auroral cycle from Korean records corresponding to the period of the 11th–18th century. For a recent review, see section 2 in Arlt & Vaquero (2020).

In a benchmark article, Eddy (1976) established that the level of solar activity in the period 1645-1715 was very low. This period was called Maunder Minimum (hereafter, MM). Cullen (1980) questioned the low level of solar activity during the MM using NESO from oriental chronicles. Clark & Stephenson (1980) strongly criticized this result and, months later, a thorough study of the seasonal variability of NESO was published (Willis et al. 1980) concluding that oriental sunspot records likely provide circumstantial evidence of solar activity. Very few years later, Xu & Jiang (1982) used the argument of the existence of NESO in this period recorded in local Chinese histories to refute the existence of the MM, but this idea was quickly rebutted by Eddy (1982).

History repeats itself again and some works in the recent past used NESO recorded during the MM to suggest high solar activity in this period (Ogurtsov et al. 2003, Zolotova & Ponyavin 2015). Recently, Wang & Li (2020) have studied sizes of NESO recorded in the period 1819–1918 from sunspot number index. Extrapolating the statistical results to the twelve NESO recorded during the MM in Yau & Stephenson (1988), these authors concluded that at least eight of those twelve NESO should be identified as large sunspots and four as giant. Based on that argument, Wang & Li (2020) suggest that the solar activity level during the MM is still an open question.

In this work, we analyze the number of groups and the sizes of the largest groups observed in days when NESO were recorded in the 19th century as Wang & Li (2020). Data used in our analysis are presented in Section 2. The analysis and discussion of these data are exposed in Section 3 and Section 4 includes the main conclusions of this work. According to our analysis, the very low level of solar activity during MM cannot be questioned using our current knowledge of NESO.

**2. Data**

There are 31 NESO recorded between 1819 and 1885 according to Wang & Li (2020). Table 1 includes some information on these NESO of the 19th century. We note that some of them seem to be the same sunspot observed in close days. One example of this fact could be the NESO recorded on 23 and 24 February 1873. In this case, we can group all NESO in 27 different "events".

We have consulted several documentary sources in order to obtain the data presented in Table 1. Thus, we have extracted useful information from the observations by Tevel (1819), Schwabe (1829-1865), Secchi (1872), the Ogyalla and Haynald observatories in Hungary (1872-1885), and the Royal Greenwich Observatory (1874-1885). We highlight the sunspot observations analyzed in this work were made by outstanding observers of that time and therefore we can consider they are of good quality. We could not find information for the following dates: 3 September 1819, 24 April 1829, 25 December 1851, 19 January 1852, 8 February 1856, 19 March 1863, and 23-24 February 1873. Instead, we provide information corresponding to the closest dates available for these dates, that is: 4 September 1819, 23 April 1829, 22 December 1851, 18 January 1852, 9 February 1856, 20 March 1863, and 19 February 1873.

Table 1. Information on solar activity in days when NESO were recorded for the period 1819–1885. Year, month, and day are represented in the first three columns, number of groups observed in that day (or the closest day) are reported in the fourth column, and the astronomer/observatory responsible of the observations ("T" is referred to Tevel, "S" to Schwabe, "SC" to Secchi, "O" to Ogyalla Observatory, GPR to Royal Greenwich Observatory and "H" to Haynald Observatory) is indicated in the sixth column. Symbol "*" depicts when there are no available records in a day when NESO were recorded and then information corresponding to the closest day is provided.

| Year | Month | Day | Groups | Observer |
|------|-------|-----|--------|----------|
| 1819 | 9 | 3 | 3 | T* |
| 1829 | 4 | 24 | 7 | S* |
| 1829 | 4 | 27 | 8 | S |
| 1839 | 8 | 14 | 5 | S |
| 1848 | 5 | 9 | 6 | S |
| 1851 | 12 | 25 | 8 | S* |
| 1852 | 1 | 19 | 5 | S* |
| 1852 | 3 | 22 | 5 | S |
| 1852 | 4 | 2 | 8 | S |
| 1852 | 12 | 29 | 6 | S |
| 1853 | 2 | 22 | 4 | S |
| 1853 | 5 | 11 | 2 | S |
| 1853 | 5 | 17 | 3 | S |
| 1853 | 6 | 21 | 3 | S |
| 1853 | 8 | 19 | 3 | S |
| 1855 | 1 | 20 | 2 | S |
| 1856 | 2 | 8 | 2 | S* |
| 1856 | 9 | 13 | 1 | S |
| 1860 | 12 | 4 | 10 | S |
| 1861 | 3 | 30 | 9 | S |
| 1861 | 11 | 24 | 5 | S |
| 1863 | 3 | 19 | 2 | S* |
| 1863 | 4 | 2 | 3 | S |
| 1865 | 4 | 9 | 1 | S |
| 1865 | 7 | 18 | 2 | S |
| 1872 | 2 | 3 | 8 | SC |
| 1873 | 2 | 23 | 8 | O* |
| 1873 | 2 | 24 | 8 | O* |
| 1874 | 12 | 9 | 3 | GPR |
| 1883 | 12 | 26 | 10 | O and H |
| 1885 | 7 | 5 | 4 | O and H |

## 3. Analysis and discussion on NESO recorded for the period 1819–1885

We have analyzed the number of groups recorded for those days corresponding to NESO included in Wang & Li (2020) during the period 1819–1885. Figure 1 depicts the number of cases (purple) and different events (green) as a function of the number of groups observed in dates when NESO were recorded. The highest values in the number of groups recorded regarding the number of individual cases are when 3 and 8 sunspot groups were observed while it is 3 sunspot groups taking into account different events. Regarding different events, the second highest value is when 8 sunspot groups were recorded. Furthermore, only around 39 % of NESO cases are recorded when the number of sunspot groups is higher than five and 42 % when less than four groups were observed. If we consider the number of different events for the period 1819–1885, the results are very similar: around 37 % of NESO events occurred when more than five groups were recorded and 63 % when the number of groups observed was five o lower (41 % when three or less groups were recorded). Thus, this first analysis shows that NESO records do not imply that solar activity in days when these NESO were reported was high.

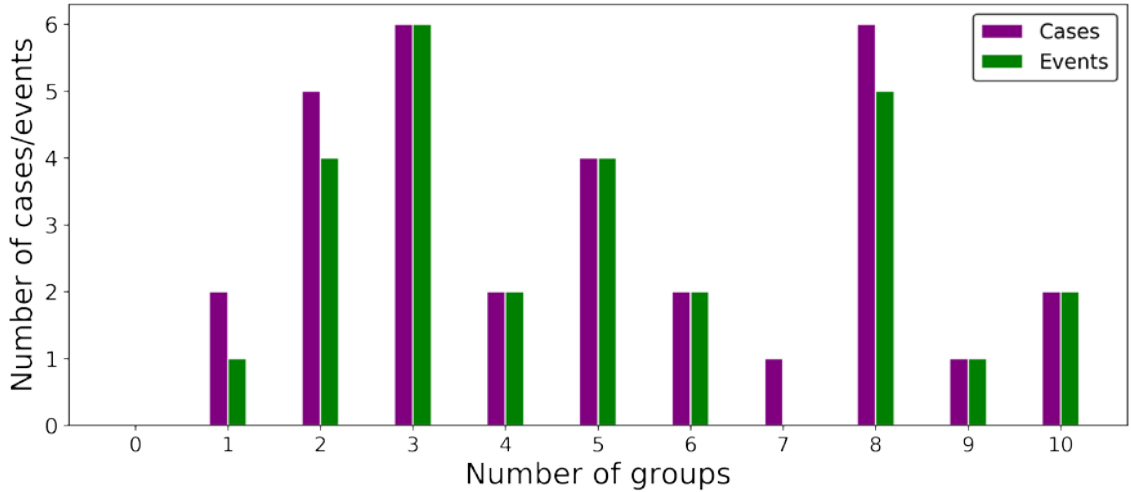

Figure 1. Number of NESO cases and different NESO events recorded during the period 1819–1885 as function of the number of sunspot groups observed in days when NESO were recorded.

We have also calculated the sizes of the largest groups recorded in days when NESO were reported during the period 1819–1885. Full spot areas (umbra and penumbra) were measured by using *Soonspot*, a software created to calculate the sunspot positions and areas (Galaviz et al. 2020). We note that: i) Schwabe's drawings corresponding to 23

April 1829 and 27 April 1829 do not present penumbral areas (Arlt et al. 2013), ii) area values corresponding to the largest groups recorded on 9 December 1874, 26 December 1883, and 5 July 1885 were taken from the website http://fenyi.solarobs.csfk.mta.hu/en/databases/GPR/, and iii) no sunspot drawing could be found corresponding to 3 February 1872 or in close dates and therefore areas were not calculated for that day. In order to analyze the area in NESO records, we used the uncorrected areas by foreshortening, which are more suitable to their study. Some works (Schaefer 1993; Vaquero & Vázquez 2009; Usoskin et al. 2015) affirm no sunspots smaller than around 500 msd can be observed by naked eye and any sunspot greater than around 1000 msd are observable. However, Neuhäuser et al. (2020) show that a good naked eye sunspot observer under determined atmospheric conditions could even detect sunspots with areas below around 150 msd. In our study, 3.2 % of areas of the largest groups recorded in days when NESO were reported were below 200 msd, 12.9 % between 200 and 499, 35.5 % between 501 and 1000 and 48.4 % above 1000 msd. This means that NESO were generally recorded in days when the area of the largest group was big but a significant percentage were recorded when areas were small (16.1 % for groups smaller than 500 msd). An example of this fact can be seen in Figure 2 where it is represented the sunspot drawing made by Schwabe on 13 September 1856 (date when one NESO was recorded) with only a small sunspot group. Therefore, NESO records do also not imply big spots necessarily. We acknowledge the difficulties to obtain area measurements from historical sunspot drawings. For example, Schwabe draw sunspots into circles of about 5 cm diameter. Thus, pores would need to have diameters of 0.04–0.1 mm and given the width of a pencil tip, at least small spots could be overestimated in the drawings (Senthamizh Pavai et al. 2015). We note that measurements calculated by Senthamizh Pavai et al. (2015) were not used in this work because only umbral areas, and not full spot areas, are provided. Moreover, the typically exaggerated spot sizes in drawings provide us upper limits and the real areas, which were seen by naked eye, are even smaller.

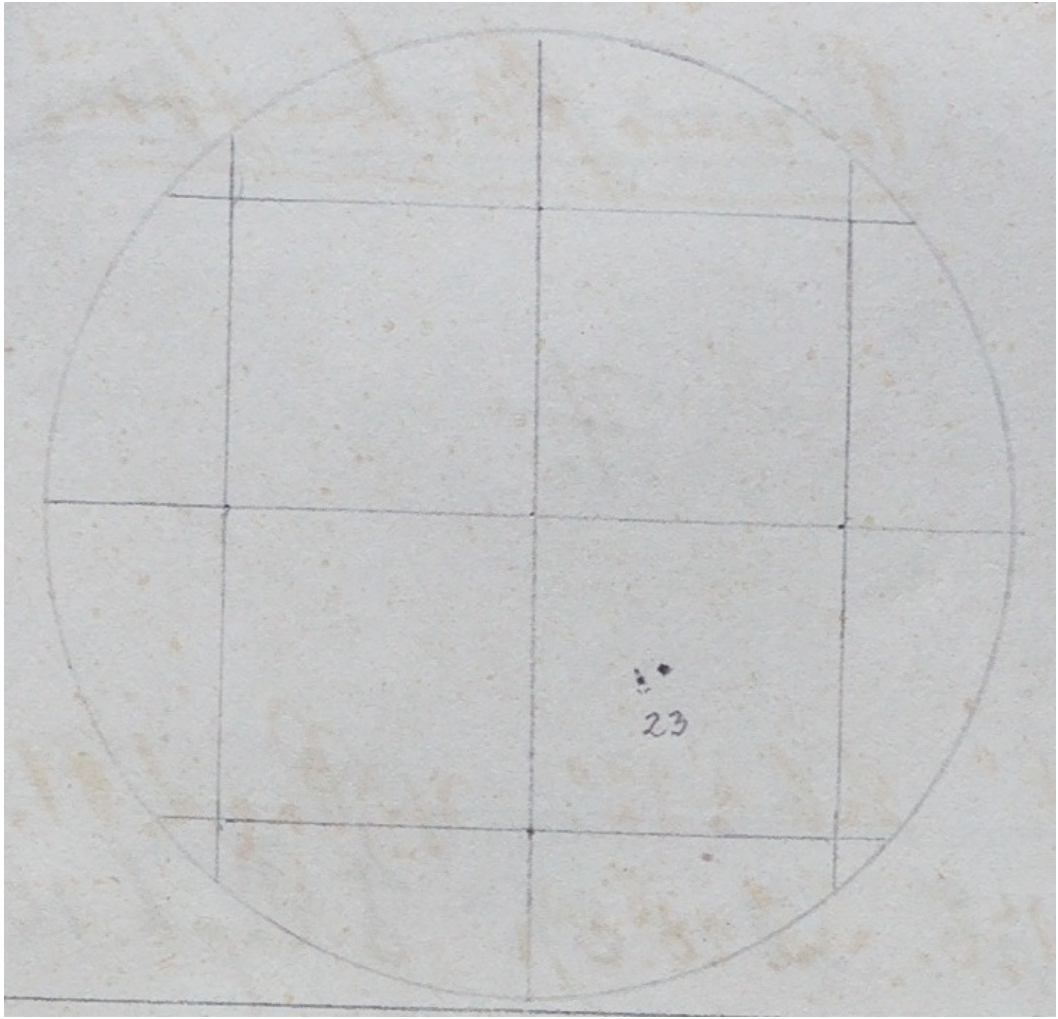

Figure 2. Sunspot drawing made by S.H. Schwabe on 13 September 1856 [RAS MS Schwabe, courtesy of the Royal Astronomical Society].

Wang & Li (2020) suggested that eight of the twelve NESO recorded during the MM in Yau & Stephenson (1988) could be large sunspots and four giant sunspots and therefore the level of solar activity for this period requires wider investigations. However, they are actually an argument for low solar activity because there are very few NESO during the core and the end of the MM. Wittmann (1992) and Xu et al. (2000) compiled more complete NESO catalogues based on the sunspot data published by Wittmann & Xu (1987) and Yau & Stephenson (1988). For example, Wittmann (1992) provides 15 NESO records during the Maunder Minimum, three more cases than Yau & Stephenson (28 July 1647, 22 May 1660, and 12 June 1659) and Xu et al. (2000) one more additional NESO record on 5 July 1661. Furthermore, regarding the number of independent NESO events (as explained previously) recorded from 1600, Wittmann (1992) and Xu et al. (2000) recorded respectively: four and three independent NESO events during the decade of the 1600s, six in both cases in the 1610s, nine and eight both in the 1620s and the 1630s, four and five in the 1640s, four in both cases in the 1650s, five and six in the 1660s, zero in both cases in the 1670s, and one both cases in the 1680s (Figure 3). Moreover, Wittmann (1992) provides zero independent NESO events in the 1690s, one in the 1700s and zero in the 1710s. In particular, if we consider the period 1667–1708, i.e. 42 years (almost all the MM), only one NESO is included in Wittmann (1992). This fact is outstanding and there is no similar example during the telescopic era regarding the same dataset by Wittmann (1992) from 1600 to 1992. In addition, we highlight that a new study of Asian dust storm activity (Chen et al. 2020) provides new arguments to understand the "high" number of NESO during the first years of the MM. Obviously, a high activity of Asian dust storms implies higher probability of NESO because the atmospheric aerosol acts as a natural filter and enlarges the time windows suitable for this kind of observation. The results obtained by Chen et al. (2020) shows a high Asian dust storm activity when some sunspots were observed at the onset of the MM. Moreover, the number of NESO records during the 18th century was also low although slightly higher than that in the MM. The main reason of this fact is probably that this century matches with a minimum in the number of dust storms in Asia (Chen et al. 2020), where the vast majority of NESO records were recorded. Then, a greater increase in the NESO records can be seen in the 19th century. We highlight that the peak of NESO records in the 1640s, 1650s, and 1660s is because a significant increase in the number of NESO records made in western civilizations with respect to other decades: 18 records in the 1840s, 13 in the 1850s, and 9 in the 1860s.

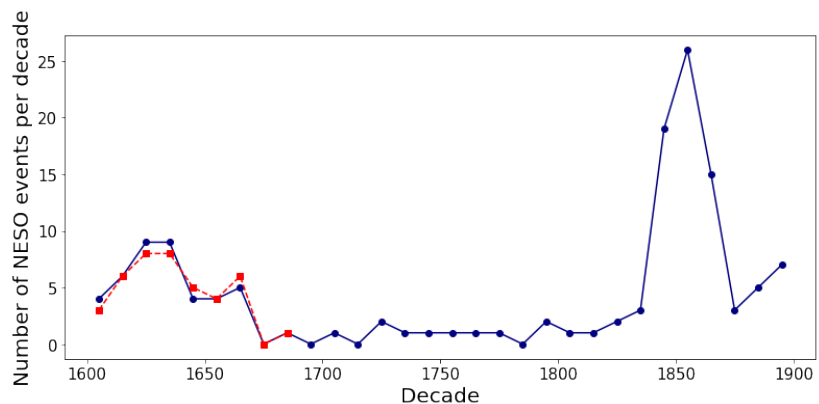

Figure 3. Number of different NESO events recorded by decade for the period 1600–1890 according to Wittmann (1992) (blue circles) and Xu et al. (2000) (red squares).

We also show an example when NESO was recorded just before the MM and only small groups were on the Sun. This is the case corresponding to 2 July 1643 when one NESO was recorded: "Within the Sun there was a black vapour shaped like a flying bird" (Yau & Stephenson 1988). However, the sunspot observation made by Hevelius (Hevelius 1647, Carrasco et al. 2019) shows two small sunspot groups for that date (Figure 4): the biggest around 300 msd (next to the western limb) and the other around below 200 msd. We also found some example of possible inconsistencies between the telescopic observations and NESO catalogues at the beginning of the MM. According to Xu (1983) including in Wittmann (1992) and Xu et al. (2000), "On the Sun there was a black light shimmering" on 12 June 1659 and "Black vapour on the Sun" was recorded on 22 May 1660, both from China. We found sunspot observations recorded by Hevelius in days close to these dates (Carrasco et al. 2015). Hevelius made observations on 6 and 13 June 1659 and on 21 and 27 May 1660 reporting no sunspot were on the Sun for these dates. Furthermore, according to the sunspot group database (Hoyt & Schatten 1998, Vaquero et al. 2016), Picard recorded one group from 7 to 19 May 1660 and then zero groups on 20 May 1660. If the observations really refer to sunspots, these could have been very small short-living groups only. Therefore, moderate or high solar activity seems unlikely in those days as well as the appearance of great sunspots on the solar disc.

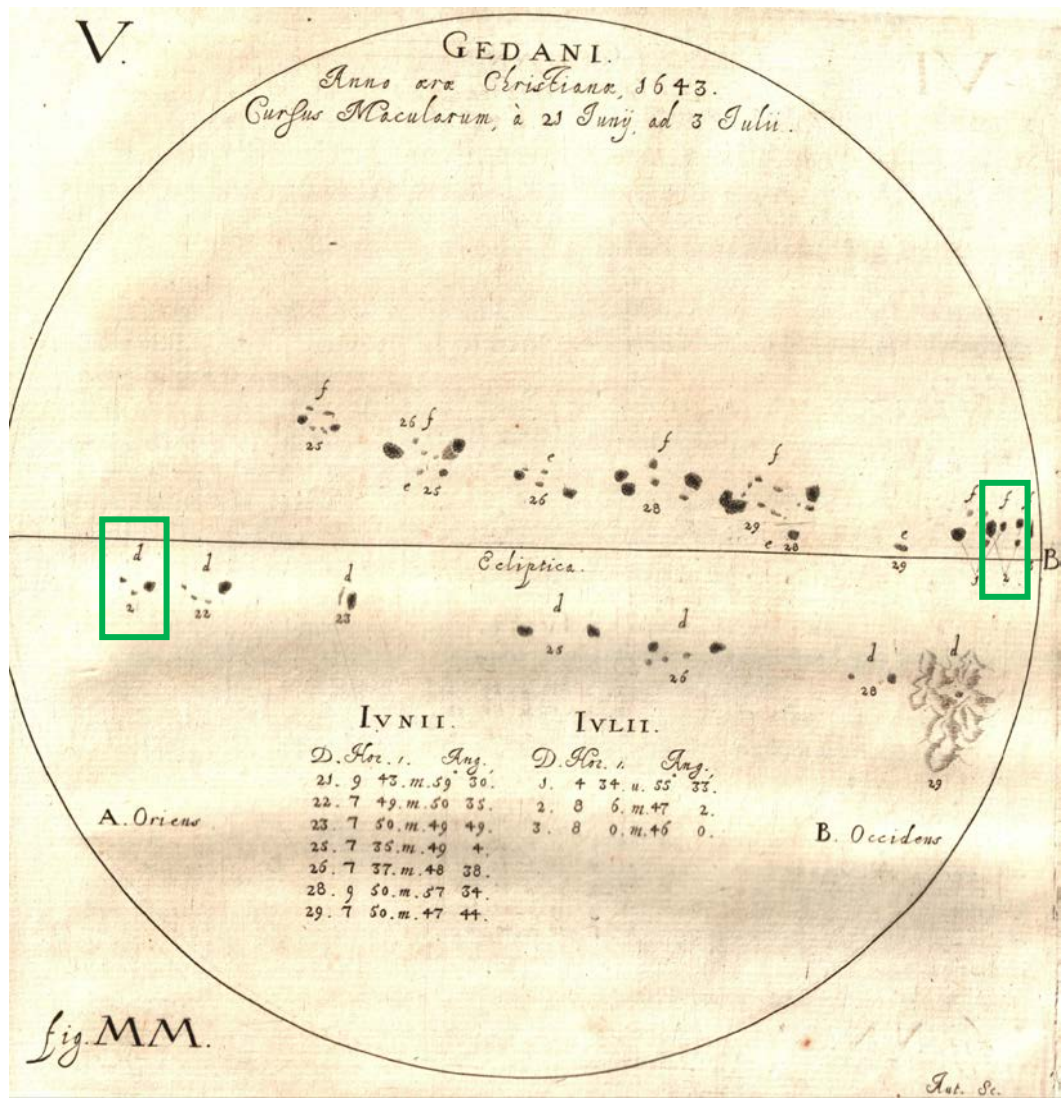

Figure 4. Sunspot drawing recorded by Hevelius in June–July 1643. The sunspot recorded on 2 July 1643 are indicated by two green squares [Source: Hevelius (1647), courtesy of the Library of the Astronomical Observatory of the Spanish Navy].

**4. Conclusions**

Wang & Li (2020) analyzed NESO records made during the period 1819–1885. Extrapolating their analysis for that period to the MM, they suggest that eight of twelve NESO recorded during the MM should be identified as large sunspots and four as giant. Wang & Li (2020) point out this fact implies that the solar activity level for the MM is still an open question despite the fact that the reconstructions based on telescopic observations show otherwise (Muñoz-Jaramillo and Vaquero, 2019).

In this work, we have studied solar activity in dates corresponding to the NESO dataset analyzed by Wang & Li (2020). We found that 42 % of NESO records analyzed in this work were recorded in days when less than four groups were observed on the Sun and only around 39 % of cases when the number of sunspot groups is higher than five. Moreover, 3.2 % of the uncorrected areas of the largest groups recorded in days when NESO were reported were below 200 msd and 12.9 % between 200 and 499 msd. Therefore, we conclude that NESO records do not necessarily imply high solar activity and big sunspot groups. We note even very low solar activity do not necessarily imply small spots (see e.g. Neuhäuser et al. 2018). We also highlight that Usoskin et al. (2015) analyzed the NESO recorded by Schwabe for the period 1825–1867 showing that around 25 % of these NESO were reported for the years when group sunspot number was below 20 and about 20 % of NESO by Schwabe were reported on days with only one group on the Sun. Thus, our results agree with those obtained by Usoskin et al. (2015) and NESO records should not be used in a simple way as an indicator of solar activity.

Regarding the MM, there is only one NESO record from 1667 to 1708 (42 years, almost all the MM). This argument could be used to show the low solar activity during the MM because there is no other similar case in the telescopic era. Furthermore, we have shown other example just before the MM from Hevelius' drawing in one day (3 July 1643) when a NESO was reported and low solar activity and only small groups were recorded. We also found some examples of possible inconsistencies between the telescopic observations and NESO records at the beginning of the MM. NESO were recorded on 12 June 1659 and 22 May 1660 from China. However, Hevelius recorded that no sunspots were observed on the Sun on 6 and 13 June 1659 and on 21 and 27 May 1660 and Picard recorded one group from 7 to 19 May 1660 and then zero groups on 20 may 1660. Thus, high solar activity and big groups seems unlikely for these days.

Therefore, our conclusions disagree with the claim by Wang & Li (2020) that the NESO indicate higher activity during the MM than usually adopted. It is also important to understand the limitations of the NESO records. Historical and meteorological factor can influence clearly this record. For example, Hameed & Gong (1991) indicated that the annual number of NESO correlates badly with 14C from 1620 probably due to historical reasons (during this epoch local historical sources were used). The role of Asian dust-storm activity can be other key factor to understand the NESO record because a higher number of dust storm, such as Chen et al. (2020) for the MM, implies higher probability of NESO. Moreover, spot size measurements in historical drawings often lead to overestimated areas. If the true areas were smaller, the fraction of small spots seen by the naked eye will be even higher. Thus, we want to emphasize that, although NESO records provide interesting information about solar activity in the past, the correct interpretation of this kind of records must be made with caution and careful taking into account the remarkable historical, sociological and climatological characteristics that also affect to NESO records.

**Acknowledgements**

This research was supported by the Economy and Infrastructure Counselling of the Junta of Extremadura through project IB16127 and grants GR18081 (co-financed by the European Regional Development Fund) and by the Ministerio de Economía y Competitividad of the Spanish Government (CGL2017-87917-P). The authors have benefited from the participation in the ISSI workshops led by M.J. Owens and F. Clette on the calibration of the sunspot number.

**Disclosure of Potential Conflicts of Interest and Ethical Statement**

The authors declare that they have no conflicts of interest. All authors contributed to the study conception and design. The first draft of the manuscript was compiled by V.M.S. Carrasco and J.M. Vaquero and all authors read and approved the final manuscript.